# SUPPRESSION OF THE MAGNETIC PHASE TRANSITION IN MANGANITES CLOSE TO THE METAL-INSULATOR CROSSOVER

F. Rivadulla[1*], J. Rivas[2], J. B. Goodenough[3]

*[1]Physical Chemistry and [2]Applied Physics Departments, University of Santiago de Compostela*

*15782-Santiago de Compostela (Spain).*

*[3]Texas Materials Institute, ETC 9.102, The University of Texas at Austin, Austin, TX 78712 (USA).*

We report the suppression of the magnetic phase transition in $La_{1-x}Ca_xMnO_3$ close to the localized-to-itinerant electronic transition, *i.e.* at $x \approx 0.2$ and $x \approx 0.5$. A new crossover temperature $T_f$ can be defined for these compositions instead of $T_C$. Unlike in common continuous magnetic phase transition the susceptibility does not diverge at $T_f$ and a spontaneous magnetization cannot be defined below it. We propose that the proximity to the doping-induced metal-insulator transition introduces a random field which breaks up the electronic/magnetic homogeneity of the system and explains these effects.

[*] For questions or comments, please contact F.R. at qffran@usc.es

After 50 years of intense research in mixed-valence manganites, the intrinsic character of the magnetic/electronic phase-segregation (PS) in these materials is widely accepted [1,2]. Moreover, it is now clear the relevance of the PS phenomenon to other doped Mott insulators, like cuprate superconductors [3], as well as classic ferromagnetic semiconductors [4].

However, a complete understanding of the thermodynamic origin of the segregated phase and its fundamental properties are still challenging our knowledge of the subject. For example, the PS phase in manganites shares some similarities with classical spin-glasses [5], but it is under discussion whether the "glassy phase" (GP) of the manganites constitutes a thermodynamically different state. The definition of the freezing temperature $T_f$ is controversial even in canonical spin-glasses, in the sense that it is unclear whether this represents a different universality class of phase transition or if it is a phase transition at all. So, a complete study of the GP in manganites should begin with the characterization of the crossover towards this phase from the high temperature paramagnetic regime. On the other hand the crucial problem of the charge density homogeneity in the PS is not solved. There are theoretical models [6,7,8] and experiments [9] which support the existence of nanometric charge density fluctuations associated to the magnetic/electronic phase separation. However, directly imaged [10,11] micrometric-size inhomogeneities seem to be incompatible with this view and extended the image of a charge homogeneous PS state, in which the two phases are only distinguishable in the magnetic/orbital arrangement [12]. In this case, the reduction in the coulombic energy would allow the dynamic coexistence of micrometer-size clusters, and a first-order phase change from polaronic to itinerant charge carriers would occur on cooling through $T_C$.

In this paper we address some of these fundamental open questions. We have applied a simple criterion based on the Landau theory to study the nature of the magnetic phase transition across the ferromagnetic-metallic (FMM) compositional range in $La_{1-x}Ca_xMnO_3$ $0.2 \leq x \leq 0.5$. Close to the localized limit, i.e. $x \approx 0.2$, $x \approx 0.5$, the first-order magnetic transition reported around $x \approx 3/8$ [13,14] is suppressed, and the system does not undergo a true magnetic phase transition. A new crossover temperature, $T_f$, is introduced by analogy with the spin-glasses. The susceptibility $\chi(H=0,T)$ does not diverge at $T_f$ and the spontaneous magnetization cannot be defined for this inhomogeneous state.

The samples were synthesized by conventional solid-state reaction from high purity reagents and the oxygen content was determined by TGA. For initial magnetization curves, the sample was heated well above $T_C$ before cooling in zero field conditions with a correction of the remanent field in the SQUID.

The inverse susceptibility in the Landau theory for magnetic phase transitions is given by [15]

$$\chi_T^{-1} = \left( \partial^2 A \big/ \partial M^2 \right)_T = a(T) + b(T)M^2 + c(T)M^4 + ... \qquad (1)$$

where $A$(T, M) is the thermodynamic Helmholtz potential and $a$(T), $b$(T), etc. are coefficients that can themselves be expanded about $T_C$ under a series of restrictions determined by the nature of the magnetic system. The convexity of $A$(T,M) with respect to M makes these coefficients necessarily positive above $T_C$, for the transition to be continuous. *An inspection of the sign of the slope of the isotherms of H/M vs. $M^2$ will then give the nature of the phase transition: positive for second order and negative for first order*. This criterion, originally proposed by Banerjee [16] was already successfully applied to determine the change in the character of the phase transition in $La_{2/3}(Ca_{1-x}Sr_x)_{1/3}MnO_3$ by Mira *et al.*[13]. Fig. 1 presents the corresponding isotherms for some representative compositions of the series between x = 0.2 and x = 0.5. For every value of x, the M vs. T curves show a similar rise of the magnetization as the temperature is lowered and the M(H) isotherms were measured around a certain temperature determined from the minimum in the ∂M/∂T, measured at low field.

For compositions close to the optimum doping level x ≈ 3/8 the M(H) isotherms present a negative slope, and hence the magnetic phase transition is first order, in agreement with previous reports [13,14]. On the other hand, the slope of the curves becomes progressively positive when the hole-density moves away from the optimal doping for $T_C$ and approaches the localized limit. Although following the Banerjee criterion this would correspond to a continuous, second order magnetic phase transition, we will argument here that the magnetic transition out of the range 0.275 < x < 0.43 is not a true phase transition, but only a change in the relative volume fractions of the fluctuations that compete to develop below a certain temperature, $T_f$.

In ordinary second-order magnetic phase transitions, the critical magnetization exponent, $\beta$, is obtained from the thermal variation of the spontaneous magnetization $M_S$(H=0,T). The values of $M_S$ at each temperature are usually derived from an extrapolation to the $M^2$ axis in the Arrot plot [17]. However, for compositions out of the first order range in Fig. 1, the isotherms never intercept the $M^2$ axis, even at temperatures much lower than the temperature of the minimum in $\partial M/\partial T$ ($T_f$). The extrapolation from low field, where the approximation is justified, neither cuts the $M^2$ axis. This makes impossible to define the order parameter. Moreover, the isotherms never reach the origin; they intercept the H/M axis at a finite value, giving a susceptibility which never diverges and hence a conventional magnetic phase transition and a true $T_C$ cannot be defined.

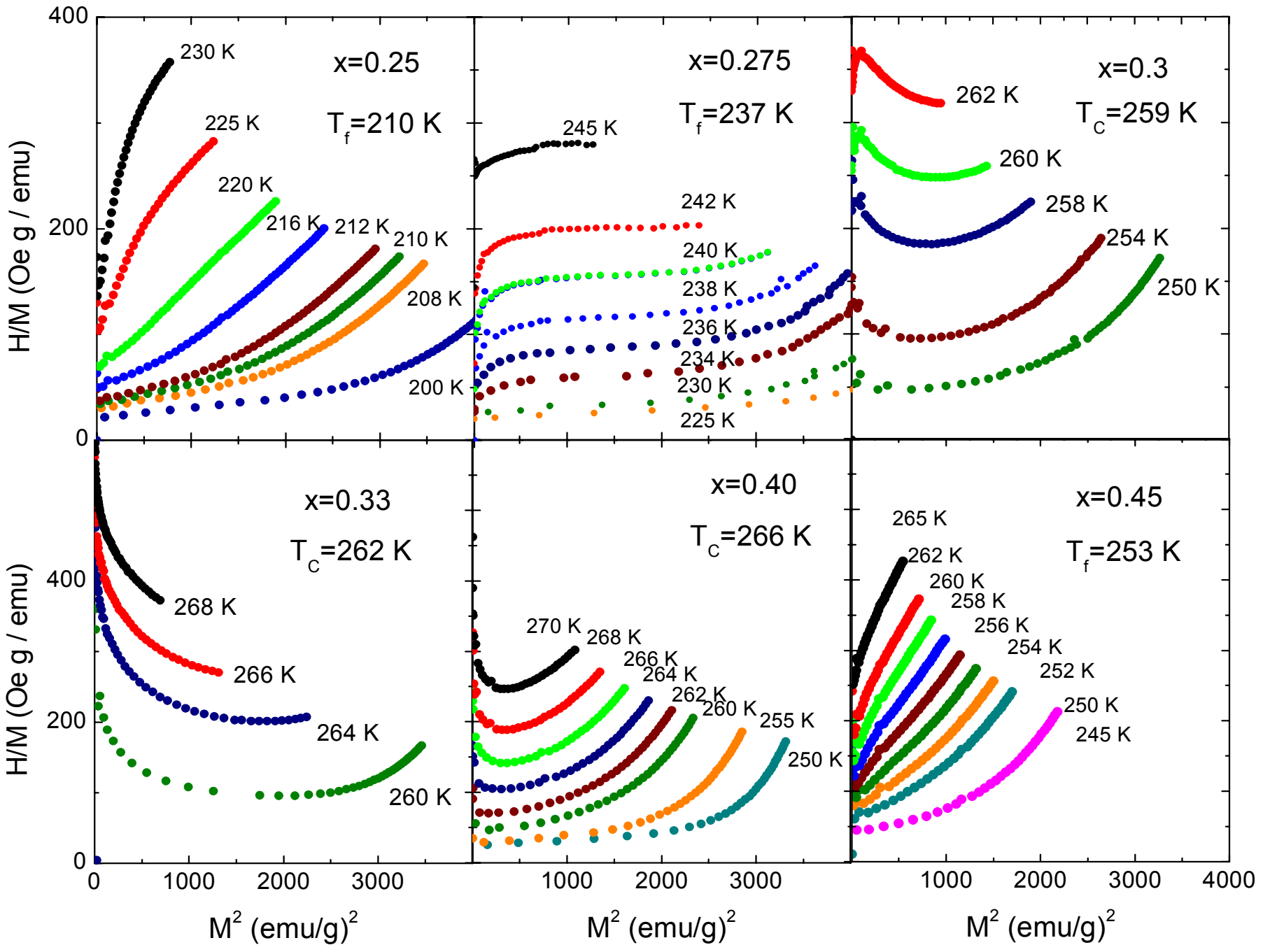

Figure 1: H/M vs. $M^2$ plots for various representative compositions around $T_C$ and $T_f$ showing the change in the sing of the slope as the composition moves through the FMM range. $T_f$ is defined instead of $T_C$, for the compositions where a conventional magnetic phase transition cannot be defined, from the minimum in the $\partial M/\partial T$ measured at low field.

In one of the few studies performed apart from x ≈ 3/8, Kim *et al* [18] reported a continuous transition with tricritical point exponents for x = 0.4; this composition was considered a borderline that separates first (x < 0.4) from second-order (x ≥ 0.4) magnetic phase transitions. However, the results reported in Fig. 1 show unequivocally that the transition *is not* first-order for all the FMM samples with x < 0.4. We considered the possibility of another tricritical point at x = 0.275 but the value of the critical exponent δ once the susceptibility was corrected, was unphysical. Moreover, there is no continuous transition beyond x = 0.4 or below x = 0.275. From this point of view, the x=0.275 and x=0.43 compositions resemble the critical point of a liquid-gas like transition. In this case the critical points are approached along a coexistence curve where two PM + FM phases are equally stable.

The suppression of the phase transition was already predicted by Aharony and Pytte [19] in models with *random fields*, and it has been observed experimentally in amorphous rare-earth alloys [20] and quenched ferrofluids with dipolar interactions [21]. In the Aharony-Pytte model the M(H) isotherms were shown to never reach the $M^2$ axis, presenting a finite susceptibility, as in our case. Imry and Ma [22] demonstrated that in 2D, a local random perturbation will break the FM system in domains of a certain size L, even when the random field is much weaker than *J*. Recently, Burgy *et al*. [23] extended the critical dimension to 3 in manganites by considering the cooperative nature of the lattice distortions in these materials. Disorder is introduced in these models as a fluctuation of *J* and *t* around the clean limit value, due to random chemical replacements in the rare earth position of the manganite.

We propose here a more general mechanism, in which the random field is introduced by the fluctuations in the magnetic/orbital ordering due to the proximity to the localized transition. By analogy to a liquid-gas transition, the thermodynamic basis of the phenomenon can be understood, without considering the chemical disorder. Due to the first order nature of the localized to itinerant electronic transition [24], the free energy, ΔG, vs. $<n>$ curve will present a double minimum with similar energies at $<n>_l$ and $<n>_i$, corresponding to the hole concentrations for the localized and itinerant regimes. The inflection points of this curve define the *spinodes*, where $\partial^2 G/\partial <n>^2 = 0$. On cooling down an initially homogeneous sample having an $<n>$ inside the spinodal region, will be unstable with respect to small fluctuations in the electronic density, giving hole-rich and hole-poor regions. Those fluctuations decrease the total free energy as the slope of ΔG vs. $<n>$ decreases past the inflection point. The Coulombic energy and the spontaneous charge-transfer between these phases will keep the system in a dynamic regime, or ordered in the form of a charge-density-wave or stripes, that are mobile unless pined by the structure. Due to the different magnetic/orbital structures of the localized and itinerant phases, these spontaneous fluctuations will introduce a random field in the system that will break it up in clusters. The suppression of the lattice thermal conductivity close to x = 0.2 and x = 0.5 [25] corroborates the dynamic coexistence of localized and itinerant clusters with different magnetic/orbital arrangement.

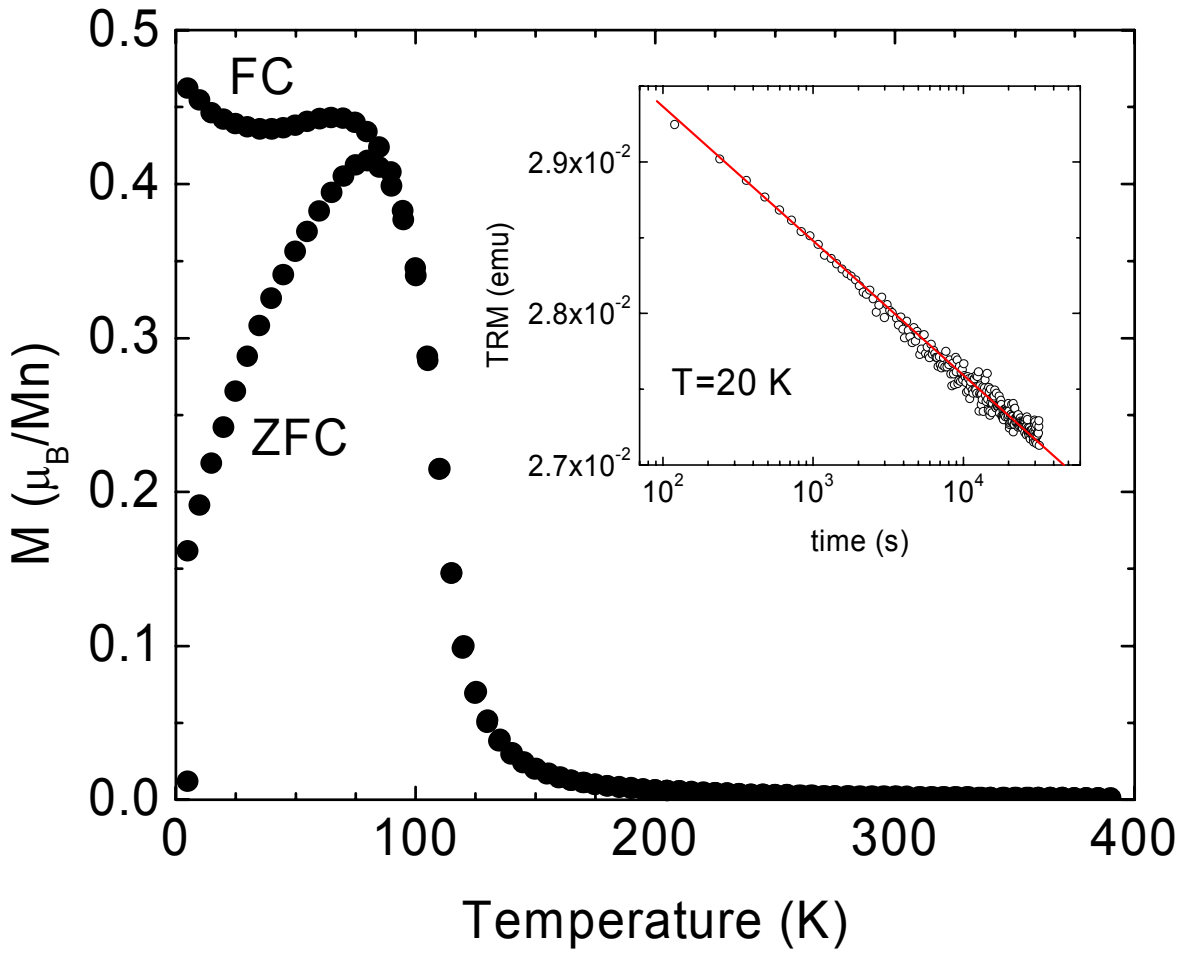

Figure 2: ZFC-FC curves for $(La_{0.25}Nd_{0.75})_{0.7}Ca_{0.3}MnO_3$ at 250 Oe ($T_f$ = 110 K). Inset: Thermoremanent magnetization (TMR) relaxation. The sample was cooled from 300 K to 20 K in 100 Oe and then the field was switched off. The data follow logarithmic time dependence. From $10^2$ to $10^4$ s the TMR diminishes only by 7%, indicating an extremely long relaxation time, compatible with FM clusters of dozens of nanometers in size.

From this point of view, the non divergence of the susceptibility at $T_f$ can be understood from the influence of finite size effects on the spin-correlation function [26]. In a real system the correlation length $\xi$ is limited by the system size L, and $\chi$(H=0,T) will saturate when $\xi$ becomes comparable to L: strictly speaking, no phase transition can be defined for a finite system at T≠0, as $\xi$ never reaches the infinity. However, finite size effects are normally negligible in macroscopic systems and only produce a rounded up of the transition very close to $T_C$. But if $\xi$ is limited like in this case to a few dozens of nanometers by the size of the FM clusters, the phase transition can be completely suppressed due to the local character of $\xi$ at every temperature [27].

To further test this hypothesis and to discard any effect of doping, we have reduced the tolerance factor of the sample with x = 0.3 introducing Nd to push it towards the localized limit. The temperature dependence of the zero-field cooling (ZFC) in $(La,Nd)_{0.7}Ca_{0.3}MnO_3$ (still in the bad-metal behavior regime, see Fig. 2), is characteristic of systems composed of random magnetic clusters with frustrated interactions; they are incompatible with a long-range FM state. The logarithmic relaxation of the thermoremanence (inset Fig. 2) is quite definitive on support of this hypothesis. The antibonding character of the $e_g$ electrons makes the volume of the localized AF fluctuations larger than the FM phase. This introduces a lattice distortion and a strain field that propagates to long range like $1/r^3$, just as a dipolar interaction. Lottis *et al.* [28] demonstrated that the decay of the magnetization of a spin system with this kind of interaction follows a logarithmic time dependence, exactly what we have observed experimentally.

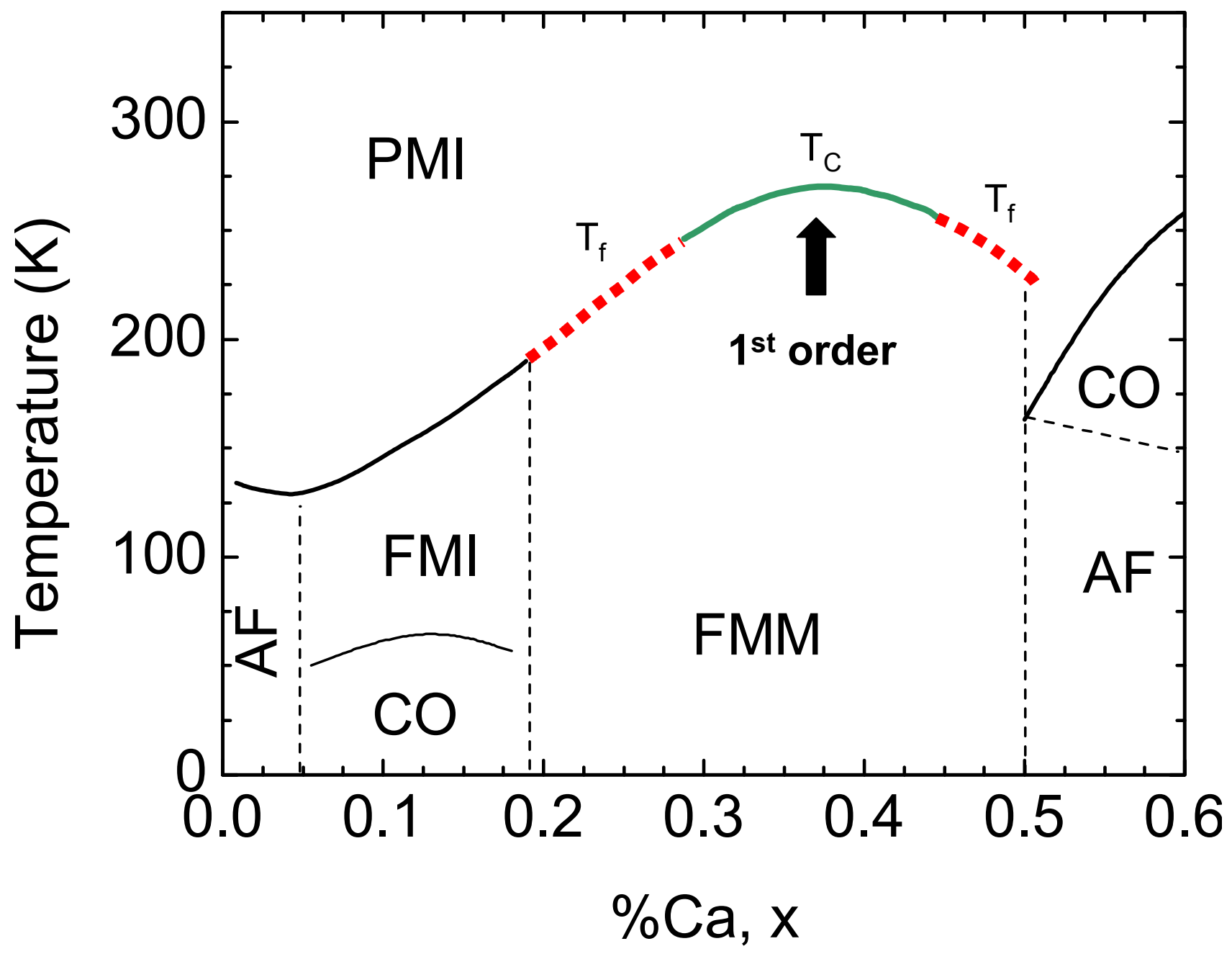

Figure 3: Revised phase diagram of $La_{1-x}Ca_xMnO_3$, (adapted from reference [29]). The dotted line indicates the $T_f$ in the regions in which a true phase transition does not occur. An inhomogeneous magnetic state with no long-range FM order develops below $T_f$.

A mobile boundary between the fluctuating phases will be created by cooperative bond-length fluctuations (BLF), as originally proposed by Goodenough [2]. In fact, BLF have been identified as the possible cause for the characteristic $\rho(T)\propto T^{3/2}$ observed in several strongly correlated metals close to the localized limit [30]. *Bad-metal* behavior and the absence of a Drude component in the optical-conductivity of the metallic-like region are also characteristic of the *vibronic* character of the mobile charges strongly coupled to the lattice fluctuations.

In figure 3 we show the revised phase diagram for the system $La_{1-x}Ca_xMnO_3$ close to the low temperature FMM range.

We also mention that a dynamic electronic phase segregation at the Mott-Hubbard transition such as that described above will lead to a continuous transfer of spectral weight between the Hubbard and the itinerant-electron bands, with the concomitant reduction in the density of states at $E_F$, opening a *pseudogap*. This pseudogap mechanism should be generally present in systems at the Mott-Hubbard transition [31] although specific characteristics of each system (screening and hybridization which will affect the Coulombic energy) will modify the extent of this phenomenon. Pseudogap features were predicted by Moreo *et al*. [32] in manganites and found in layered $La_{1.2}Sr_{1.8}Mn_2O_7$ by Dessau *et al*. [33]. The existence of a pseudogap in high-$T_C$ superconductors close to the Mott transition is also well known [34].

In summary, our results show that the magnetic phase transition in $La_{1-x}Ca_xMnO_3$ is suppressed close to the localized limit. A spontaneous fluctuation between phases with small differences in <n> and exchange constants introduces a random field that suppresses the first order character of the magnetic transition in this system. This dynamic phase-segregation phenomenon is quite general, applicable to the copper-oxide superconductors as well as the CMR manganites. We point out that the electronic inhomogeneity provides a mechanism for the formation of a pseudogap in systems close to the Mott-Hubbard transition.

**Acknowledgments.**

We acknowledge fruitful discussion with some of the participants of the "Imagine Magnetic and Superconducting Materials" workshop in Barcelona, October 2003, especially J. Fernández-Rossier, N. D. Mathur, E. Dagotto, D. Khomskii, C. A. Ramos, L. E. Hueso, J. Mira and J. Castro. We also acknowledge financial support from Ministery of Science and Technology of Spain (FEDER MAT2001-3749).

---

[1] E. Dagotto in *Nanoscale Phase Separation and Colossal Magnetoresistance*, Springer Series in Solid-State Science, Vol. 136, 2003.

[2] P. M. Raccah, J. B. Goodenough, J. Appl. Phys. **39**, 1209 (1968); J. B. Goodenough in *Structure and Bonding,* Vol. **98**, Ed. by J. B. Goodenough, Springer, Berlín, 2001.

[3] E. Dagotto, Rev. Mod. Phys. **66**, 763-840 (1994)

[4] E. L. Nagaev, Phys. Rep. **346**, 387 (2001).

[5] I. G. Deac, S. V. Diaz, B. G. Kim, S.-W. Cheong, and P. Schiffer, Phys. Rev. B **65**, 174426 (2002); See also Chapter 13 in reference [1] by P. Schiffer.

[6] S. Yunoki, Phys. Rev. Lett. **80**, 845 (1998).

[7] A. Moreo, S. Yunoki, E. Dagotto, Science **283**, 2034 (1999).

[8] F. Guinea, G. Gómez-Santos, D. P. Arovas, Phys. Rev. B **62**, 391 (2000); D. P. Arovas, G. Gómez-Santos, F. Guinea, Phys. Rev. B **59**, 13569 (1999).

[9] M. Hennion, F. Moussa, G. Biotteau, J. Rodriguez-Carvajal, L. Pinsard, A. Revcolevschi, Phys. Rev. Lett. **81**, 1957 (1998).